\documentclass[aps,twocolumn,prl,fixfloat]{revtex4-2}
\usepackage{graphicx,amsmath,amsfonts,bm, color}

\usepackage{xcolor,hyperref}
\hypersetup{
   colorlinks,
   linkcolor={blue!50!black},
   citecolor={blue!50!black},
   urlcolor={blue!80!black}
}

\usepackage{enumitem}

\newcommand{\1}{^{\mbox{\tiny (1)}}}

\newcommand{\dbar}{{\,\mathchar'26\mkern-12mu d}}
\DeclareMathAlphabet{\mathitbf}{OML}{cmm}{b}{it}

\begin{document}

\title{Internally-stressed and positionally-disordered minimal complexes \\ yield glasslike nonphononic excitations}
\author{Avraham Moriel}
\affiliation{Chemical \& Biological Physics Department, Weizmann Institute of Science, Rehovot 7610001, Israel}

\begin{abstract}
Glasses, unlike their crystalline counterparts, exhibit low-frequency nonphononic excitations whose frequencies $\omega$ follow a universal $\mathcal{D}\!\left(\omega\right)\!\sim\!\omega^4$ density of states. The process of glass formation generates positional disorder intertwined with mechanical frustration, posing fundamental challenges in understanding the origins of glassy nonphononic excitations. Here we suggest that \emph{minimal complexes} --- mechanically-frustrated and positionally-disordered local structures --- embody the minimal physical ingredients needed to generate glasslike excitations. We investigate the individual effects of mechanical frustration and positional disorder on the vibrational spectrum of isolated minimal complexes, and demonstrate that ensembles of marginally stable minimal complexes yield $\mathcal{D}\!\left(\omega\right)\!\sim\!\omega^4$. Furthermore, glasslike excitation emerge by embedding a single minimal complex within a perfect lattice. Consequently, minimal complexes offer a conceptual framework to understand glasslike excitations from first principles, as well as a practical computational method for introducing them into solids.
\end{abstract}

\maketitle
\emph{Introduction.}--- Understanding the low-frequency spectrum of crystalline materials allows calculations of thermal conductance, scattering coefficients, and various other material properties~\cite{kittel1976,landau1980statistical}. In contrast, the origin of glassy nonphononic low-frequency excitations~\cite{Maloney2004b,Maloney2006,Shimada2018}, whose corresponding frequencies $\omega$ follow a universal $\mathcal{D}\!\left(\omega\right)\!\sim\!\omega^4$ vibrational density of states~\cite{Lerner2016b,Mizuno2017,Kapteijns2018,Lopez2020,Bonfanti2020,Lerner2018}, is not fully understood. Understanding these excitations from first principles is of prime importance as they govern glasses' physical properties such as heat transport~\cite{Kittel1949,Flubacher1959,Zeller1971}, scattering~\cite{Hehlen2000,Ruffle2003,Monaco2009,Monaco2009a,Zanatta2010,Rossi2011,Chumakov2014,Moriel2019,Kapteijns2020a}, and plastic response~\cite{Maloney2004b,Maloney2006,Guo2007,Kumar2013,Deschamps2014,Ketkaew2018,Shimada2018}.

Glasses are typically formed via rapidly quenching a liquid~\cite{Cavagna2009}. During this process, self-organization leads to positional disorder intertwined with mechanical frustration and local variations in elastic stiffness~\cite{Alexander1998,Cavagna2009}. These generic properties motivated previous investigations of the effects of preparation protocol, variations of composition, and internal stress's amplitude on glassy excitations~\cite{Lerner2017,Kapteijns2018a,Rainone2020,Gonzalez2020,Lerner2018,Moriel2019,Kapteijns2020a}. Others exploited structural measures~\cite{Srolovitz1981,Cubuk2015,Schoenholz2017,Richard2020} to probe the glass transition and irreversible processes, both related to the emergence of glassy excitations. However, the origins of these excitations remain obscure because glass's positional disorder and mechanical frustration are inseparable.

Theoretical approaches capturing glassy features usually avoid the explicit treatment of positional disorder and mechanical frustration, either by coarse-graining procedures (e.g. in mean-field approaches~\cite{Alexander1998,Wyart2005,DeGiuli2014a,Kirkpatrick1988,Parisi2010a,Lin2015,Benetti2018}), or by relying on assumptions regarding statistical microscopic properties (e.g. random-matrix methods~\cite{Beltukov2011,Manning2015,Stanifer2018}). Other approaches~\cite{Buchenau1992,Gurevich2002,Gurevich2003,Parshin2007} \emph{a priori} assume the existence of specific localized structures to predict $\mathcal{D}\!\left(\omega\right)\!\sim\!\omega^4$. While informative on their own, these approaches evade discussing what minimal elementary ingredients generate glassy excitations.

In this Letter we show how both mechanical frustration and positional disorder generate glasslike low-frequency excitations from first principles. We study the harmonic behavior of ordered minimal systems --- \emph{minimal complexes}. We then examine the distinct roles of mechanical frustration and positional disorder by independently introducing them to minimal complexes. Coupling mechanical frustration and positional disorder by approaching mechanical instability, minimal complexes yield glasslike excitations and $\mathcal{D}\!\left(\omega\right)\!\sim\!\omega^4$. Overall, the approach taken here unveils the roles played by positional disorder and mechanical frustration, offers a minimal analytical model for understanding glasslike excitations and a glassy length-scale, and presents a practical method for introducing these excitations into solids.

\emph{Minimal complexes.}--- Both mechanical frustration and positional disorder modify the harmonic vibrational spectrum of a solid. To demonstrate the effect of mechanical frustration, consider a system of $N$ particles in $\dbar$ spatial dimensions, of total potential energy $U$, under force balance $\frac{\partial U}{\partial \bm{x}}\!=\!\bm{0}$ (where $\bm{x}$ is a $\dbar N$-dimensional position vector). When the system is stable, the Hessian $\bm{\mathcal{M}}\!\equiv\!\frac{\partial^2 U}{\partial \bm{x} \partial \bm{x}}$ is positive semi-definite, and its eigenmodes $\bm{\psi}$ and corresponding frequencies $\omega$ govern the system's harmonic vibrational spectrum via the eigenvalue equation $\bm{\mathcal{M}}\cdot\bm{\psi}\!=\!\omega^2\bm{\psi}$ (masses taken to unity).

For simplicity, let us focus on pairwise interactions of the form $\varphi_\alpha\!\equiv\!\varphi\left(\Delta_\alpha\right)$, with  $\bm{\Delta}_{\alpha}\!\equiv\!\left(\bm{x}_j-\bm{x}_i\right)^T$ being the difference vector of the $\alpha^{\text{th}}$'s bond (here $\bm{x}_i$ is the position of the $i^{\text{th}}$ particle), and $\Delta_{\alpha}\!\equiv\!\left|\bm{\Delta}_{\alpha}\right|$ its magnitude. The Hessian $\bm{\mathcal{M}}$ may be further decomposed as~\cite{Alexander1998,Wyart2005,Lerner2018}
\begin{equation}\label{eq:hessian_decomp}
  \bm{\mathcal{M}} = \bm{\mathcal{H}} + \bm{\mathcal{F}} \ ,
\end{equation}
where the elastic stiffnesses $\varphi_\alpha''\!\equiv\!\tfrac{\partial^2 \varphi_\alpha}{\partial \Delta_\alpha \partial \Delta_\alpha}$ of all interactions contribute to $\bm{\mathcal{H}}$, and the internal stresses $\varphi'_\alpha\!\equiv\!\tfrac{\partial \varphi_\alpha}{\partial \Delta_\alpha}$ from all interactions contribute to $\bm{\mathcal{F}}$~\cite{SM}.

The force configuration satisfying the force balance condition $\frac{\partial U}{\partial \bm{x}}\!=\!\bm{0}$ has major implications on the resulting $\bm{\mathcal{M}}$ and its vibrational spectrum. For stress-free systems in which \emph{each} interaction within the system contributes zero force, $\varphi_\alpha'\!=\!\bm{0}$, $\bm{\mathcal{F}}\!=\!\bm{0}$, and the classical harmonic approximation $\bm{\mathcal{M}}\!=\!\bm{\mathcal{H}}$ holds~\cite{landau1976mechanics,SM}. However, systems in which only the \emph{net force} vanishes for each particle, internal stresses exist resulting in a sizable $\bm{\mathcal{F}}$  --- internal stresses modify the harmonic vibrational spectrum even under force balance~\cite{Alexander1998,Wyart2005,Lerner2018}.

Let us first examine how $\bm{\mathcal{H}}$ and $\bm{\mathcal{F}}$ contribute to the single bond Hessian $\bm{\mathcal{M}}_{1}\!=\!\tfrac{\partial^2\varphi_1}{\partial \bm{x}\partial \bm{x}}$. In general, $\bm{\mathcal{M}}_{1}$ is $2\dbar$-dimensional, containing $2\dbar$ eigenvalues $\lambda$'s and eigenmodes $\bm{\psi}$'s. Translational invariance yields $\dbar$ zero-modes. The single bond stiffness $\bm{\mathcal{H}}_{1}\!\propto\!\varphi_1'' \hat{\bm{\Delta}}_1 \hat{\bm{\Delta}}_1^T$ contributes a single eigenvalue $\lambda\!\propto\!\varphi''_{1}$ and a corresponding $\bm{\psi}$ along $\hat{\bm{\Delta}}_{1}$~\cite{SM}. The single bond $\bm{\mathcal{F}}_{1}\!\propto\!\frac{\varphi_1'}{\Delta_1} \left(\bm{\mathcal{I}}_{\dbar} - \hat{\bm{\Delta}}_1 \hat{\bm{\Delta}}_1^T\right)$ ($\bm{\mathcal{I}}_{\dbar}$ being the $\dbar$-dimensional identity matrix) contributes the remaining $\dbar - 1$ eigenvalues $\lambda\!\propto\!\frac{\varphi'_{1}}{\Delta_{1}}$ corresponding to $\bm{\psi}$'s orthogonal to $\hat{\bm{\Delta}}_{1}$~\cite{SM}. The presence of internal stresses alters the resulting vibrational spectrum, and may even destabilize $\bm{\mathcal{M}}_{1}$ once $\varphi'_{1}\!<\!0$.

\begin{figure}[t]
\centering
\includegraphics[width=0.5\textwidth]{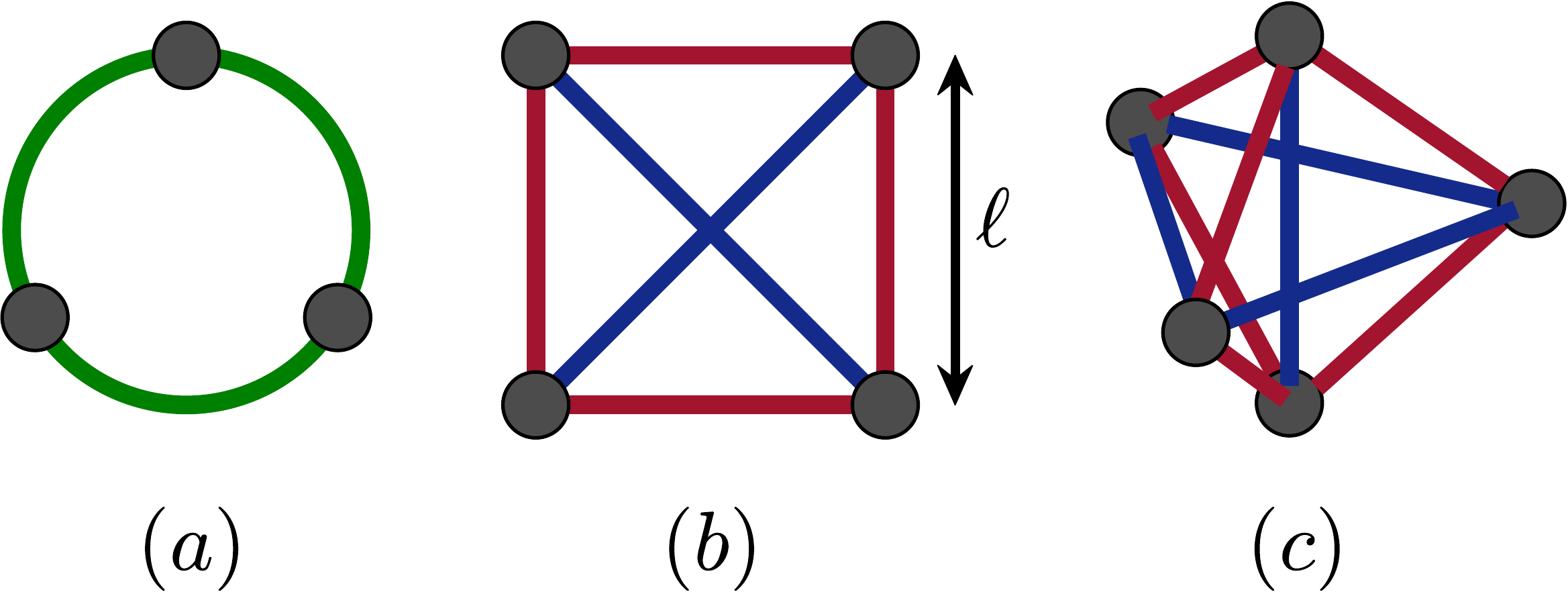}
\caption{Examples of minimal complexes in (a) $\dbar\!=\!1$, (b) $2$ and (c) $3$. In (a) forces are $1$-dimensional, and each particle has two interactions. Force balance is satisfied by choosing a constant force of any magnitude. In (b), bonds are colored according to a force-balanced internally stressed state --- red bonds are repulsive, and blue bonds are attractive (this could be reversed by a negative multiplication scaling factor). In (c) a similar coloring scheme is used.}
\label{fig:fig1}
\end{figure}

The configuration of internal stresses, if present, must satisfy $\frac{\partial U}{\partial \bm{x}}\!=\!\bm{0}$. What minimal system allows such a configuration in the first place? A single particle under force balance imposes $\dbar$ constraints. Interaction with $\dbar + 1$ neighbors ensures the existence of non-trivial solutions to these $\dbar$ equations~\cite{Alexander1998}. A \emph{minimal complex} --- a fully-connected system with a minimal number of $\dbar+2$ particles [and $(\dbar+1)(\dbar+2)/2$ interactions] --- ensures the existence of a \emph{single} internally-stressed force-balanced state (also known as a state of self stress~\cite{Lubensky2015,Mao2017,SM}). Figure~\ref{fig:fig1} shows possible realizations of minimal complexes in $\dbar\!=\!1$, $2$ and $3$.

Consider the Hessian $\bm{\mathcal{M}}_{mc}$ of an internally-stressed minimal complex. For simplicity, we consider a minimal complex in $\dbar\!=\!2$ consisting of 4 particles arranged in a perfect square of side-length $\ell$, and 6 interactions [cf. Fig.~\ref{fig:fig1}(b)]. We focus on $\dbar\!=\!2$ as it is the minimal spatial dimension required for non-trivial internal stress contributions, and choose a square geometry both because of its symmetry, and as it may serve as a simple unit-cell in a 2D lattice. First, we set all stiffness to a constant $\varphi''_{\alpha}\!=\!\kappa$ to highlight the role of internal stresses. Then, we find the \emph{single} allowed configuration of forces that produces zero net force on all the particles and impose such internal stresses multiplied by the amplitude $\xi$~\cite{Lubensky2015,Mao2017,SM}. We set $\xi\!>\!0$ to correspond to short-range repulsion and long-range attraction, while $\xi\!<\!0$ corresponds to the opposite case.

We normalize the minimal complex's Hessian $\hat{\bm{\mathcal{M}}}_{mc}\!\equiv\!\bm{\mathcal{M}}_{mc}/\kappa$, and decompose it according to $\hat{\bm{\mathcal{M}}}_{mc}\!=\! \hat{\bm{\mathcal{H}}}_{mc} + \epsilon \hat{\bm{\mathcal{F}}}_{mc}$,  with $\hat{\bm{\mathcal{H}}}_{mc}\!\equiv\!\bm{\mathcal{H}}_{mc}/\kappa$,  $\hat{\bm{\mathcal{F}}}_{mc}\!\equiv\!\bm{\mathcal{F}}_{mc}\ell/\xi$, and $\epsilon\!\equiv\!\xi/\kappa \ell$ capturing the importance of internal stresses relative to elastic forces. We drop the $\bullet_{mc}$ subscript for readability, and use $\hat{\bm{\mathcal{H}}}$, $\hat{\bm{\mathcal{F}}}$ and $\hat{\bm{\mathcal{M}}}$ for the minimal complex quantities exclusively.

As both $\hat{\bm{\mathcal{H}}}$ and $\hat{\bm{\mathcal{F}}}$ are translationally and rotationally invariant, both have three zero eigenvalues $\lambda\!=\!0$.
\onecolumngrid\

\begin{figure*}[b!]
\centering
\includegraphics[width=\textwidth]{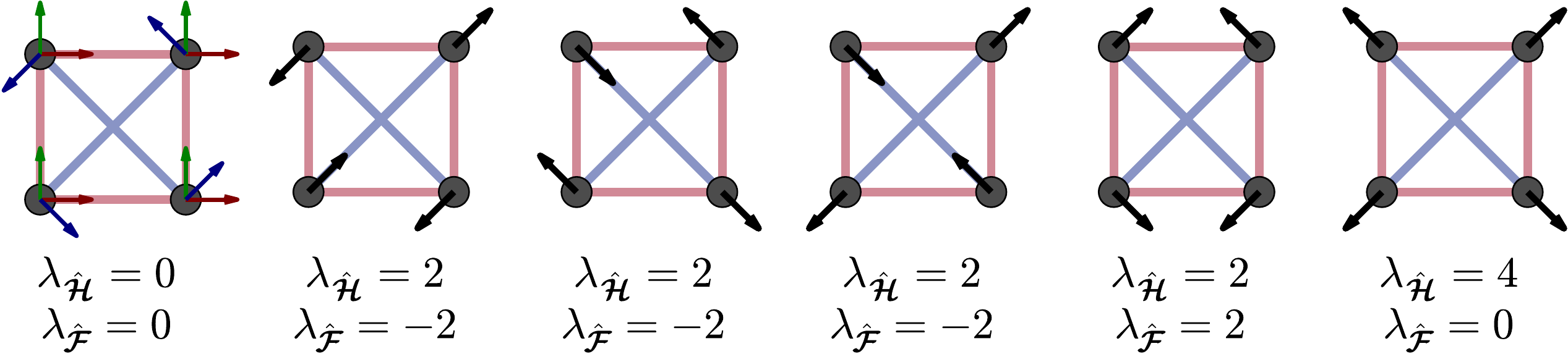}
\caption{Visualization of the eigenmodes $\bm{\psi}$'s and eigenvalues $\lambda$'s of $\hat{\bm{\mathcal{H}}}$ and $\hat{\bm{\mathcal{F}}}$. The three translational and rotational zero-modes of $\hat{\bm{\mathcal{H}}}$ and $\hat{\bm{\mathcal{F}}}$ are shown in red, green, and blue respectively (left). The four-fold degeneracy of the $\lambda_{\hat{\bm{\mathcal{H}}}}\!=\!2$ shear mode band is lifted in the presence internal stresses $\hat{\bm{\mathcal{F}}}$ to a three-fold degenerate band, and a single mode. The three-fold degenerate band is characterized by shearing of repulsive (red) interactions, while the single mode is obtained by shearing attractive (blue) interactions. While dilation is associated with $\lambda_{\hat{\bm{\mathcal{H}}}}\!=\!4$, it is an additional zero-mode for $\hat{\bm{\mathcal{F}}}$ as no bond is sheared.}
\label{fig:fig2}
\end{figure*}
\twocolumngrid\
\\
\begin{figure}[ht]
\centering
\includegraphics[width=0.48\textwidth]{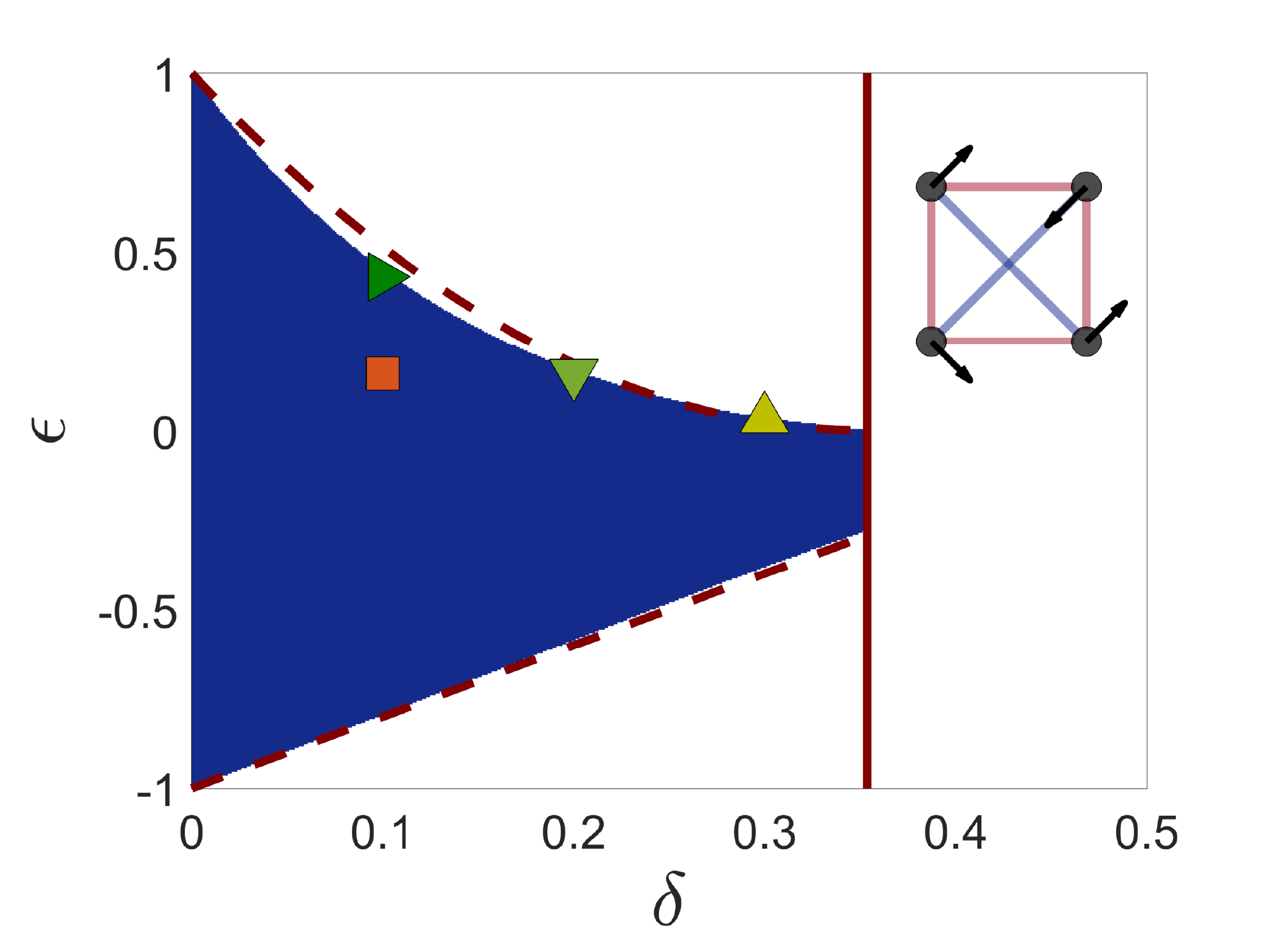}
\caption{Stability phase diagram of $\hat{\bm{\mathcal{M}}}$ as a function of the positional disorder $\delta$ and the internal stress $\epsilon$. Each point in the phase diagram corresponds to the lowest non-zero eigenvalue of $10^5$ realizations within the $\left(\delta,\epsilon\right)$ ensemble. Blue regions signify stable ensembles (positive semi-definite $\hat{\bm{\mathcal{M}}}$'s), while white regions indicate $\left(\delta,\epsilon\right)$ ensembles with at least a single unstable minimal complex. Solid and dashed red lines correspond to theoretical predictions. The inset depicts a representative destabilizing positional perturbation for the $\epsilon\!>\!0$ stability boundary. Markers correspond to the $\left(\delta,\epsilon\right)$ ensembles examined in Fig.~\ref{fig:fig4}}
\label{fig:fig3}
\end{figure}
\\
$\hat{\bm{\mathcal{H}}}$ captures a stress-free elastic system, its eigenmodes correspond to shear and dilation vibrations. Specifically, $\hat{\bm{\mathcal{H}}}$ has $4$ shear modes of $\lambda_{\hat{\bm{\mathcal{H}}}}\!=\!2$ and a single $\lambda_{\hat{\bm{\mathcal{H}}}}\!=\!4$ dilation mode. The eigenmodes of $\hat{\bm{\mathcal{F}}}$ share similar spatial form, but their associated eigenvalues $\lambda_{\hat{\bm{\mathcal{F}}}}$ differ. As $\hat{\bm{\mathcal{F}}}$ is associated with shearing motion~\cite{SM}, its dilation mode becomes a zero-mode $\lambda_{\hat{\bm{\mathcal{F}}}}\!=\!0$. Shear vibrations split into a three-fold degenerate band of $\lambda_{\hat{\bm{\mathcal{F}}}}\!=\!-2$, and a single mode $\lambda_{\hat{\bm{\mathcal{F}}}}\!=\!2$. Both $\lambda$'s and $\bm{\psi}$'s of $\hat{\bm{\mathcal{H}}}$ and $\hat{\bm{\mathcal{F}}}$ are visualized in Fig.~\ref{fig:fig2}.

The resulting spectrum of $\hat{\bm{\mathcal{M}}}$ consists of three zero-modes, a three-fold degenerate band of $\lambda_{\hat{\bm{\mathcal{M}}}}\!=\!2\left(1-\epsilon\right)$ shear modes, a single shear mode $\lambda_{\hat{\bm{\mathcal{M}}}}\!=\!2\left(1+\epsilon\right)$, and a dilation mode $\lambda_{\hat{\bm{\mathcal{M}}}}\!=\!4$. The presence of internal stresses breaks the symmetry of shear deformation, and lifts the degeneracy in $\hat{\bm{\mathcal{M}}}$'s spectrum. In fact, $\hat{\bm{\mathcal{M}}}$'s positive semi-definiteness is ensured only when internal stresses are small compared to the elastic forces, $|\epsilon|\!<\!1$.

As mentioned above, glasses exhibit positional disorder in addition to internal stresses. To explore the role played by positional disorder, we introduce a random perturbation vector $A\left(\cos\left(\theta_i\right),\sin\left(\theta_i\right)\right)^T$ of amplitude $A$ to the position of each particle in the minimal complex (the $4$ angles $\theta_i\!\in\!\left.[0,2\pi\right.)$ drawn from a uniform distribution). The dimensionless parameter $\delta\equiv A/\ell$ captures the magnitude of the positional disorder amplitude $A$ relative to the side-length $\ell$.

Generically, positional disorder lifts the degeneracy in $\hat{\bm{\mathcal{M}}}$'s spectrum. However, each positional perturbation modifies the spectrum of $\hat{\bm{\mathcal{M}}}$ differently. To probe these differences we consider ensembles of minimal complexes characterized by $\left(\delta,\epsilon\right)$ and extract their lowest non-zero eigenvalue. A negative minimal eigenvalue implies at least a single realization is unstable; otherwise all realizations within the $\left(\delta,\epsilon\right)$ ensemble account for stable energetic minima.

The stability of the $\left(\delta,\epsilon\right)$ ensembles shown in Fig.~\ref{fig:fig3} reveals clear boundaries between stable and unstable ensembles. To understand these boundaries, consider first the $\epsilon\!<\!0$ regime, in which the destabilizing eigenvalue is $\lambda_{\hat{\bm{\mathcal{M}}}}\!=\!2\left(1+\epsilon\right)$. Linear perturbation theory predicts the lowest eigenvalue vanishes at the critical strain $\epsilon_c\!=\!2\delta-1$. This analytical prediction is plotted in Fig.~\ref{fig:fig3} and agrees with the numerical results.

In the case of $\epsilon\!>\!0$ the $\lambda_{\hat{\bm{\mathcal{M}}}}\!=\!2\left(1-\epsilon\right)$ degenerate band destabilizes $\hat{\bm{\mathcal{M}}}$. We established above that $\epsilon_c\!=\!1$ for $\delta\!=\!0$. Also, once three particles are aligned --- corresponding to $\delta\!=\!1/\sqrt{8}$ --- the system effectively reduces to a single-bond embedded in $\dbar\!=\!2$, unstable under internal stresses; beyond this point, it is unlikely the ensemble will stabilize again. Finally, due to degeneracy we assume non-negligible second order corrections in $\delta$. Altogether, we predict $\epsilon_c\!=\!8(\delta - \tfrac{1}{\sqrt{8}})^2$, as well as a critical line at $\delta\!=\!1/\sqrt{8}$ --- both in agreement with our numerical findings, as shown in Fig.~\ref{fig:fig3}.

\emph{Glasslike nonphononic excitations.}--- The two aspects discussed so far --- internal stresses and positional disorder --- are essential features of glasses~\cite{Lerner2018,Alexander1998,Wyart2005}. The process of glass formation couples between positional disorder to internal stresses --- what is an analogous coupling between these essential features in minimal complexes?

\begin{figure}[!ht]
\centering
\includegraphics[width=0.45\textwidth]{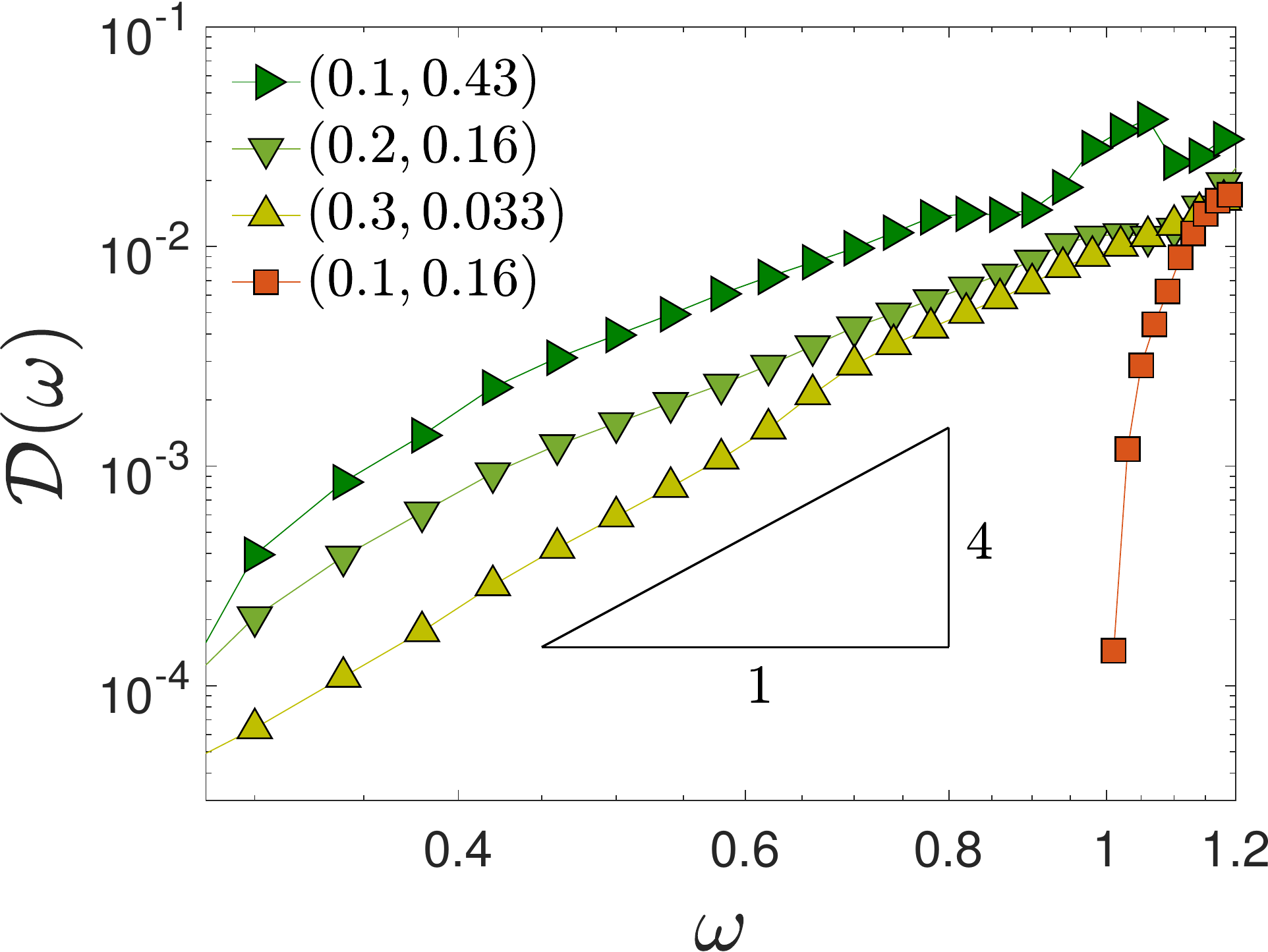}
\caption{Density of states $\mathcal{D}\left(\omega\right)$ of four different $\left(\delta,\epsilon\right)$ ensembles [using $10^7$  realizations for each $\left(\epsilon,\delta\right)$ combination], as denoted in Fig.~\ref{fig:fig3}. Marginal ensembles' low-frequency spectrum (green-yellow triangles) follow a power-law distribution close to the glassy $\mathcal{D}\left(\omega\right)\!\sim\!\omega^4$~\cite{Lerner2016b,Mizuno2017,Kapteijns2018,Lopez2020,Bonfanti2020,Lerner2018}. The non-marginal ensemble (orange square) does not exhibit such power-law scaling, emphasizing the importance of mechanical marginality as a coupling mechanism.}
\label{fig:fig4}
\end{figure}

Glassy modes are easily identified near mechanical instabilities~\cite{Maloney2004b,Maloney2006,Shimada2018}; we hypothesize minimal complexes' marginally stable ensembles would be of our interest. We treat marginal stability as an effective coupling between positional disorder $\delta$ and internal stresses $\epsilon$, and sample $\left(\delta,\epsilon\right)$ ensembles as denoted in Fig.~\ref{fig:fig3}. As conventionally short-range interactions are repulsive and long-range interactions are attractive, we confine the discussion to the $\epsilon\!>\!0$ regime. Frequencies $\omega\!\equiv\!\sqrt{\lambda}$ of marginal ensembles indeed follow $\mathcal{D}\!\left(\omega\right)\!\sim\!\omega^4$ as shown in Fig.~\ref{fig:fig4}. Coupling positional disorder with internal stresses through mechanical marginality yields a glasslike density of states.

\begin{figure}[!ht]
  \centering
  \includegraphics[width=0.5\textwidth]{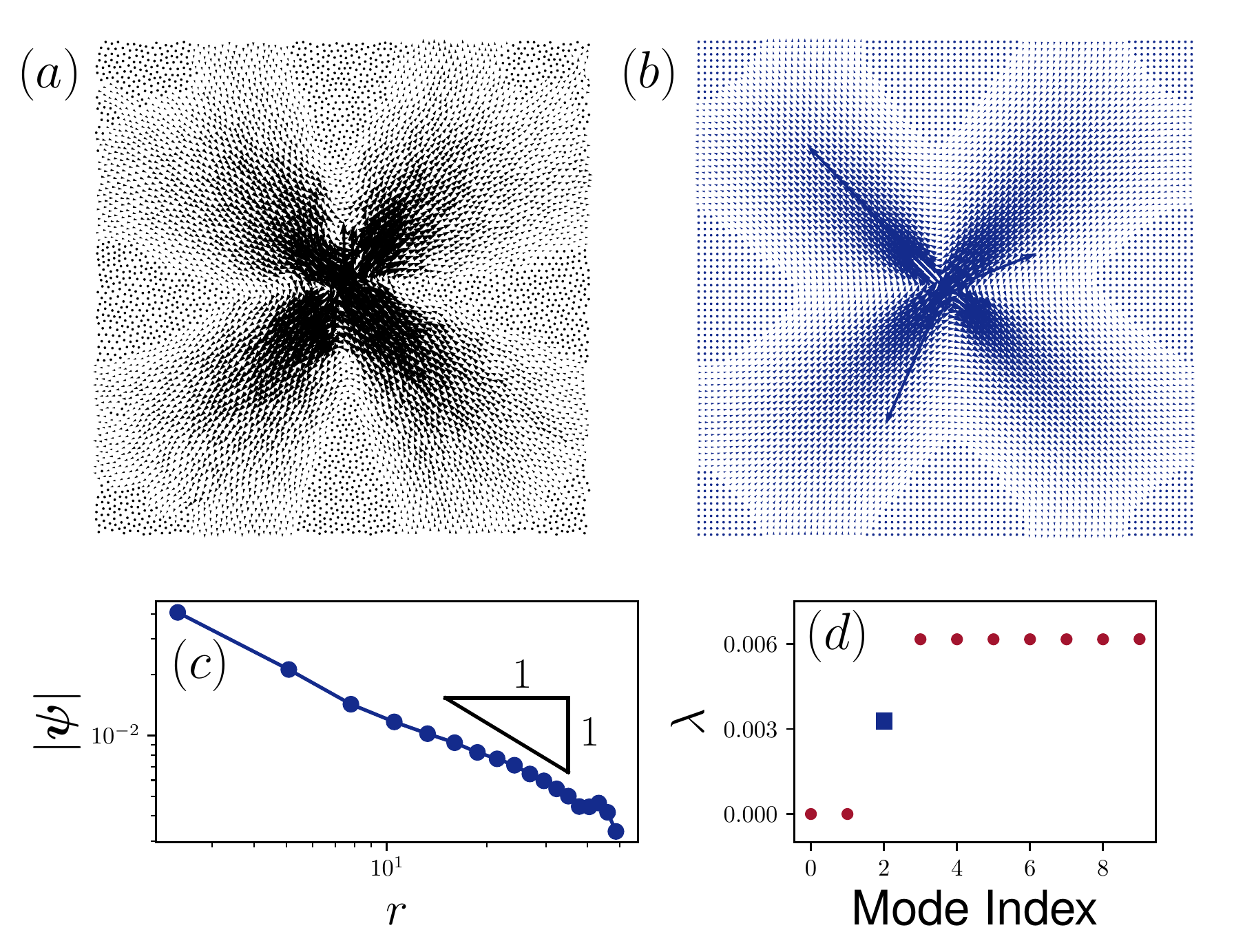}
  \caption{(a) An example of a glassy mode from an inverse-power-law glass ($N\!=\!80^2$, details in ~\cite{Moriel2020}). (b) An example of the emerging glasslike mode $\bm{\psi}$ from a minimal complex embedded within a lattice ($N\!=\!80^2$), obtained with $\delta\!=\!0.1$ and $\epsilon\!=\!1.44$ (enlarging the system increases $\epsilon_c$). The obtained quadropolar structure is reminiscent of the one observed in (a). (c) Decay of the magnitude $\left|\bm{\psi}\right|\!\equiv\!\sqrt{\bm{\psi}\cdot\bm{\psi}}$, as a function of the distance $r$ away from the core, scaling as $r^{-1}$, similar to glassy modes~\cite{Kapteijns2018,Moriel2020}. (d) The first $10$ eigenvalues of $\bm{\mathcal{M}}$ of the system. The mode presented in (b) is marked in blue, shown to exist below the first phononic band.}
  \label{fig:fig5}
\end{figure}

Glasses' $\mathcal{D}\!\left(\omega\right)\!\sim\!\omega^4$ scaling corresponds to the presence of glassy nonphononic excitations~\cite{Lerner2016b,Lerner2018}; we expect minimal complexes's low frequency excitations to exhibit similar spatial structure. To test this, we construct a stress-free lattice with a unit-cell of similar spatial structure to the minimal complex~\cite{SM} [cf. Fig.~\ref{fig:fig1}(b)]. We choose one unit-cell and introduce to it a positional perturbation $\delta$ and internal stresses $\epsilon$, essentially embedding a minimal complex within the ordered lattice. We then extract the lowest non-zero eigenmode $\bm{\psi}$ from the full system's $\bm{\mathcal{M}}$. An example of $\bm{\psi}$ is shown in Fig.~\ref{fig:fig5}. Embedding a minimal complex within an ordered, stress-free medium results in spatial features similar to those of glassy modes~\cite{Maloney2004b,Maloney2006,Shimada2018}.

\emph{Discussion.}--- In this Letter we demonstrated how coupling between positional disorder and internal stresses in minimal complexes generates two glassy characteristics: a $\mathcal{D}\!\left(\omega\right)\!\sim\!\omega^4$ glassy density of states~\cite{Lerner2016b,Mizuno2017,Kapteijns2018,Lopez2020,Bonfanti2020,Lerner2018} and glasslike nonphononic excitations~\cite{Maloney2004b,Shimada2018,Maloney2006}. Utilizing the simplicity of $\dbar\!=\!2$ minimal complexes, we analytically predicted the effects of internal stresses on the vibrational spectrum, and derived stability conditions in the presence of positional disorder. We then coupled positional disorder and internal stresses via mechanical marginality --- analogous to the self-organization dynamics during typical quenching procedures --- to generate glasslike nonphononic excitations and a glassy density of states. Overall, minimal complexes --- glasses' ``spherical cows'' --- provide transparent insights into the inner workings of glassy vibrational spectrum.

Above, we exposed the role of internal stresses in glassy physics. While internal stresses in real glasses do not localize, we demonstrated above how their presence may yield a glassy length-scale. In fact, our observations also provide a lower limit on such a length-scale~\cite{EffeDeGiuli2014a,Lerner2016b,Rainone2020} --- glasslike excitations arise for a minimum of $\dbar+2$ particles. Extending the analysis beyond pairwise interactions and to higher spatial dimensions may reveal model-specific glassy characteristics. Finally, while other mechanisms of generating glassy modes exist~\cite{Gonzalez2020}, the mechanism described above is of fundamental importance and is universal across several different classes of glassy materials, from foams to metallic glasses.

\emph{Acknowledgements.}--- We thank Talya Vaknin, Yuri Lubomirsky, Edan Lerner, and Eran Bouchbinder for insightful comments and discussions, and for critically reading the manuscript. We acknowledge support from the Minerva Foundation with funding from the Federal German Ministry for Education and Research, the Ben May Center for Chemical Theory and Computation, and the Harold Perlman Family.


%


\onecolumngrid
\newpage
\begin{center}
\textbf{\large Supplemental Materials for: \\ ``Internally-stressed and positionally-disordered minimal complexes \\ yield glasslike nonphononic excitations''}
\end{center}
\twocolumngrid
\setcounter{equation}{0}
\setcounter{figure}{0}
\setcounter{table}{0}
\setcounter{section}{0}
\setcounter{page}{1}
\makeatletter
\renewcommand{\theequation}{S\arabic{equation}}
\renewcommand{\thesection}{S-\Roman{section}}
\renewcommand{\thefigure}{S\arabic{figure}}
\renewcommand*{\thepage}{S\arabic{page}}
\renewcommand{\bibnumfmt}[1]{[S#1]}
\renewcommand{\citenumfont}[1]{S#1}

In this Supplemental Materials we provide: detailed mathematical derivation of the elastic stiffnesses $\bm{\mathcal{H}}$, and the internal stresses $\bm{\mathcal{F}}$,  mathematical procedure for obtaining the single state of internal stress, and details about embedding minimal complexes in a $\dbar\!=\!2$ ordered lattice.

\subsection{Explicit elastic stiffnesses and internal stresses contributions} \label{sse:hfm}
Consider a system of $N$ particles in $\dbar$ spatial dimensions. Assuming $N_b$ pairwise interactions $\varphi_\alpha\!\equiv\!\varphi\left(\Delta_\alpha\right)$, with  $\bm{\Delta}_{\alpha}\!\equiv\!\left(\bm{x}_j-\bm{x}_i\right)^T$ being the difference vector of the $\alpha^{\text{th}}$ bond (here $\bm{x}_i$ is the position of the $i^{\text{th}}$ particle) and $\Delta_{\alpha}\!\equiv\!\left|\bm{\Delta}_{\alpha}\right|$ its magnitude, the total potential energy $U$ of the system is cast as
\begin{equation}\label{eq:energy}
  U = \sum_{\alpha=1}^{N_b} \varphi_\alpha \ .
\end{equation}
The Hessian $\bm{\mathcal{M}}$ is then obtained as~\cite{S-Lerner2018}
\begin{equation}\label{eq:hessian}
  \mathcal{M}_{ij} = \sum_{\alpha=1}^{N_b} \Gamma^T_{i \alpha }\left[\varphi''_{\alpha} \left(\hat{\bm{\Delta}}_{\alpha}\hat{\bm{\Delta}}_{\alpha}^{T} \right) + \frac{\varphi'_{\alpha}}{\Delta_{\alpha}}\left(\bm{\mathcal{I}}_{\dbar} - \hat{\bm{\Delta}}_{\alpha}\hat{\bm{\Delta}}_{\alpha}^{T}\right)\right]\Gamma_{\alpha j}
   ,
\end{equation}
where $\Gamma_{\alpha i}\!\equiv\!\left(\delta_{ki}-\delta_{ji}\right)\otimes \bm{\mathcal{I}}_\dbar$ is the $\dbar N_b\!\times\! \dbar N$ system's (signed) incidence matrix~\cite{S-strang1993}, $\alpha$ describes the interaction between particles $j$ and $k$, $\delta_{ki}$ is the Kronecker delta, $\otimes$ denotes the Kronecker product, and $\bm{\mathcal{I}}_\dbar$ is the $\dbar$-dimensional identity matrix. Here $\hat{\bm{\Delta}}_{\alpha}\!\equiv\!\bm{\Delta}_{\alpha} / \Delta_{\alpha}$ is the normalized $\dbar$-dimensional difference vector, $\varphi'_\alpha\!\equiv\!\tfrac{\partial \varphi\left(\Delta_\alpha\right)}{\partial \Delta_\alpha}$ and $\varphi''_{\alpha}\!\equiv\!\tfrac{\partial^2 \varphi\left(\Delta_\alpha\right)}{\partial \Delta_\alpha \partial \Delta_\alpha}$.

We follow by decomposing $\bm{\mathcal{M}}$ as~\cite{S-Lerner2018}
\begin{equation}\label{eq:hessian_decomp}
  \bm{\mathcal{M}} = \bm{\mathcal{H}} + \bm{\mathcal{F}} \ ,
\end{equation}
to elastic stiffnesses $\bm{\mathcal{H}}$, and internal stresses $\bm{\mathcal{F}}$ contributions. Explicitly, $\bm{\mathcal{H}}$ and $\bm{\mathcal{F}}$ take the form
\begin{subequations} \label{eq:H_F}
\begin{align}
  \mathcal{H}_{ij} \equiv& \sum_{\alpha=1}^{N_b} \Gamma^T_{i \alpha }\varphi''_{\alpha} \left(\hat{\bm{\Delta}}_{\alpha}\hat{\bm{\Delta}}_{\alpha}^{T} \right) \Gamma_{\alpha j} \ , \label{eq:H}\\
  \mathcal{F}_{ij} \equiv& \sum_{\alpha=1}^{N_b} \Gamma^T_{i \alpha}\frac{\varphi'_{\alpha}}{\Delta_{\alpha}}\left(\bm{\mathcal{I}}_{\dbar} - \hat{\bm{\Delta}}_{\alpha}\hat{\bm{\Delta}}_{\alpha}^{T}\right)\Gamma_{\alpha j}  \ , \label{eq:F}
\end{align}
\end{subequations}
In the absence of internal stresses $\varphi'_\alpha\!=\!0$, $\bm{\mathcal{F}}$ vanishes and the classical harmonic limit $\bm{\mathcal{M}}\!=\!\bm{\mathcal{H}}$ is recovered~\cite{S-landau1976mechanics}.

A two-particle system with a single interaction has a total of $2\dbar$ degrees of freedom. Of these, $\dbar$ are zero modes corresponding to translational invariance, guaranteed by the force balance criterion $\frac{\partial U}{\partial \bm{x}}\!=\!\bm{0}$. Thus the single bond Hessian $\bm{\mathcal{M}}_{1}\!=\!\tfrac{\partial^2\varphi_1}{\partial \bm{x}\partial \bm{x}}$ has $\dbar$ non-trivial eigenmodes and eigenvalues. Of these, a single eigenvalue originates from $\bm{\mathcal{H}}_1\!\propto\!\varphi''_{1} \hat{\bm{\Delta}}_{\alpha}\hat{\bm{\Delta}}_{\alpha}^{T}$, which is a rank $1$ matrix, with a corresponding $\bm{\psi}$ along $\hat{\bm{\Delta}}_1$, and $\lambda\!\propto\!\varphi''_{1}$. Similarly, $\bm{\mathcal{F}}_1$ is a rank $\dbar-1$ matrix, contributing the remanning $\dbar-1$ eigenmodes $\bm{\psi}$'s orthogonal to $\hat{\bm{\Delta}}_1$, and $\lambda$'s $\!\propto\!\frac{\varphi'_{1}}{\Delta_{1}}$. These observations facilitate the intuition that $\bm{\mathcal{H}}$ is associated with vibrations along the bond's directions, while $\bm{\mathcal{F}}$ is associated with shearing motion.

Finally, note that while we have assumed only pairwise interactions in Eq.~\eqref{eq:energy}, it is possible to decompose three-body interactions (and higher) in a similar manner: a contribution from the second derivative of the interaction (analogous to $\varphi''_\alpha$ in the pairwise case), and a contribution from a single derivative of the interaction (analogous to $\varphi'_\alpha$ in the pairwise case). While we haven't included such terms in our analysis to maintain simplicity, these interactions did not affect the observed $\mathcal{D}\!\left(\omega\right)\!\sim\!\omega^4$ accross different classes of computer glasses~\cite{S-Richard2020}.

\subsection{Single state of internal stresses and minimal complex construction} \label{sse:isconf}
As mentioned, minimal complexes by construction allow for a \emph{single} internal stress configuration (up to a multiplicative constant) that satisfies the force balance constraint $\frac{\partial U}{\partial \bm{x}}\!=\!\bm{0}$. To see how only one such configuration emerges, note that the total degrees of freedom in a minimal complex is $\dbar N\!=\!\dbar \left(\dbar + 2\right)$, where $N_0\!=\!\dbar\left(\dbar+1\right)/2$ of them are zero modes associated with rigid body translations and rotations, and the total number of constraints is $N_b\!=(\dbar+1)(\dbar+2)/2$. The number of internal stress configurations $N_s$ (also known as states of self stress) is given by an index theorem~\cite{S-Mao2017} as $N_0+N_b-\dbar N \!=\!N_s \!=\! 1$ for minimal complexes in any spatial dimension $\dbar$.

To find this single allowed configuration we cast the force balance constraint using $\bm{\Gamma}^T$. To ensure force balance, we have to satisfy
\begin{equation}\label{eq:forceBalance}
  \sum_\alpha \Gamma^T_{i \alpha} t_\alpha \hat{\bm{\Delta}}_\alpha\!=\!\bm{0} \ ,
\end{equation}
where $t_\alpha$ is the tension in the $\alpha^{\text{th}}$ bond. This demand must be satisfied for every spatial dimension $s\!=\!1...\dbar$ and for each particle $i\!=\!1...\dbar+2$.

We now confine the discussion to 2D minimal complexes, as we used these to obtain glassy excitations and density of states. First, we position four particles at the vertexes of a square of $\ell\!=\!1$. Then, we sample $4$ angles $\theta_i\!\in\!(0,2\pi]$ from a uniform distribution, and displace each of the $4$ particles by $\bm{x}_i\!\rightarrow\! \bm{x}_i + A (\cos \theta_i, \sin \theta_i)^T$.

We construct auxiliary $N_b\!\times\!N$ incidence matrixes corresponding to spatial dimension $s$ as $\Gamma_{\alpha i}^{(s)}\!\equiv\!\left(\delta_{ki}-\delta_{ji}\right) \hat{\Delta}_{\alpha}^{(s)}$ where the indexes $j$ and $k$ correspond to the $\alpha^{\text{th}}$ bond. Then, we construct the full $N_b\!\times\!\dbar N$ geometric incidence matrix as $\bm{\Gamma}^{\dbar}\equiv\left(\bm{\Gamma}^{(1)},...,\bm{\Gamma}^{(\dbar)}\right)$ containing both connectivity information (encapsulated in the $\delta_{\bullet i}$'s terms), and positional information (encapsulated in the $\hat{\Delta}$'s). In $\dbar\!=\!2$, this matrix rakes the form $\bm{\Gamma}^{2}\equiv\left(\bm{\Gamma}^{(x)},\bm{\Gamma}^{(y)}\right)$. The null-space of $\left(\bm{\Gamma}^{\dbar}\right)^T$ then contains the single $N_b$ dimensional internal stress configuration $\bm{v}$ that ensures force balance $\left(\bm{\Gamma}^{\dbar}\right)^T \bm{v}\!=\!\bm{0}$.

After finding the null-space vector $\bm{v}$, we construct a projection matrix $\bm{P}\!\equiv\!\bm{v}\left(\bm{v}^T \bm{v}\right)^{-1}\bm{v}^T$~\cite{S-strang1993}. Next, we sample a random vector $\bm{t}^{rnd}$ of length $(\dbar+1)(\dbar+2)/2$ (its entries drawn from a uniform distribution $t^{rnd}_i\!\in\!(0,1]$), and project it using the projection matrix $\tilde{\bm{t}}\!=\!\bm{P}\cdot\bm{t}^{rnd}$. We then normalize it according to the smallest entry corresponding to the peripheral interactions $\alpha\!\in\!\bigcirc$ [where $\bigcirc$ denotes these peripheral bonds, i.e. red interactions in Fig.~1(b) in the manuscript] and multiply the resulting vector by $\xi$, $\bm{t}\!=\!\xi \tilde{\bm{t}} / (\underset{\alpha\in\bigcirc}{\text{min}} \tilde{\bm{t}})$. This procedure ensures $\xi\!>\!0$ corresponds to repulsive peripheral interactions and attractive inner interactions.

Finally, we set $\varphi_\alpha'\!=\!-t_\alpha$, and $\varphi''_\alpha\!=\!\kappa$ (we use $\kappa\!=\!1$ for convenience). We construct $\bm{\mathcal{H}}_{mc}$ and $\bm{\mathcal{F}}_{mc}$ separately, normalize them according to $\hat{\bm{\mathcal{H}}}_{mc}\!=\!\bm{\mathcal{H}}_{mc}/\kappa$, $\hat{\bm{\mathcal{F}}}_{mc}\!=\!\bm{\mathcal{F}}_{mc}\ell/\xi$, and construct $\hat{\bm{\mathcal{M}}}_{mc}\!=\!\hat{\bm{\mathcal{H}}}_{mc} + \epsilon \hat{\bm{\mathcal{F}}}_{mc}$, with $\epsilon\!=\!\xi/\kappa \ell$, as described in the main text.

We used MATLAB~\cite{S-matlab} to generate ensembles of minimal complexes characterized by $\left(\delta,\epsilon\right)$, diagonalize their respective Hessians, and obtain their density of states.

\subsection{Embedding minimal complexes in a 2D lattice} \label{sse:lattice}
We first construct a $\dbar\!=\!2$ lattice of $N$ cites, using the shape of the $\dbar\!=\!2$ minimal complex as a unit-cell. We employ periodic boundary conditions, as is conventionally used in molecular dynamics simulations~\cite{S-Deng1989}, and set all interactions' stiffness to $\varphi''_\alpha\!=\!\kappa$. Then, we choose a single unit-cell within the lattice, and perform similar procedures as described in Sec.~\ref{sse:isconf} for obtaining the internal stress allowed configuration only for this specific unit cell. We then construct $\bm{\mathcal{H}}$ and $\bm{\mathcal{F}}$ for the entire lattice, where the only contribution to $\bm{\mathcal{F}}$ originates from the internally-stressed minimal complex embedded within the lattice. We used Python~\cite{S-python} to generate such a lattice with $N=80^2$, diagonalize its Hessian, and visualize its lowest non-zero eigenmode.

In computer glasses, a similar procedure can be followed if at least a single minimal complex --- $\dbar+2$ fully-connected particles --- exists. While introducing internal stresses may be inconsistent with the used microscopic potential, it may serve as a powerful tool to measure local marginality and ``softness'', which are of interest in rheological and structural contexts~\cite{S-Cubuk2015,S-Zylberg2017,S-Richard2020}.

%

\end{document}